
\input phyzzx


\doublespace
\font\zfont = cmss10 scaled \magstep1
\def\ZZ{\hbox{\zfont Z\kern-.4emZ}}

\def\mgravitino{m_{3/2}}
\def\intD{\int d^4 \theta}

\def\mb{m_b}
\def\mlsp{m_{\rm LSP}}
\def\mLSP{m_{\rm LSP}}

\def\nlsp{N_{\rm LSP}}
\def\omegab{\Omega_b}
\def\omegalsp{\Omega_{\rm LSP}}
\def\OmegaLSP{\Omega_{\rm LSP}}
\def\Omegalsp{\Omega_{\rm LSP}}
\def\Omegab{\Omega_b}
\def\mphi{m_{\phi}}
\def\nphi{n_{\phi}}
\def\bd{\bar{d}}
\def\bu{\bar{u}}
\def\be{\bar{e}}


\frontpagetrue

\let\picnaturalsize=N
\def\picsize{1.2in}
\def\picfilename{scipp_tree.eps}

\let\nopictures=Y
\centerline{} \vskip-3pc
\ifx\nopictures Y\else{\ifx\epsfloaded Y\else\input epsf \fi
\let\epsfloaded=Y
{\line{\hbox{
\ifx\picnaturalsize N\epsfxsize \picsize\fi
{\epsfbox{\picfilename}}} \hfill
\vbox{                        
\singlespace
\hbox{SLAC-PUB-95-6917}
\vskip.6in
}
}}}\fi
\rightline{SLAC-PUB-95-6917} \vskip-14pt
\rightline{hep-ph/9506274}
\vskip10pt


\vskip.1in

\def\SLAC{\centerline{\it Stanford Linear Accelerator Center}
   \centerline{\it Stanford University}
   \centerline{\it Stanford, CA  94309} }

\vskip-.3in
\title{\seventeenbf Baryons and Dark Matter from the Late
Decay of a Supersymmetric Condensate}

\centerline{{\fourteencp Scott Thomas}
\foot{Work supported by the Department of Energy under contract
DE-AC03-76SF00515.}}
\singlespace
\SLAC

\vskip.6cm
\vbox{
\centerline{\bf Abstract}

The possibility that both the baryon asymmetry and dark matter arise
from the late decay of a population of supersymmetric
particles is considered.
If the decay takes place below the LSP freeze out temperature,
a nonthermal distribution of LSPs results.
With conserved $R$ parity these relic LSPs contribute to
the dark matter density.
A net asymmetry can exist in the population of decaying
particles if it arises from coherent production
along a supersymmetric flat direction.
The asymmetry is transferred to baryons if the condensate decays
through the lowest order nonrenormalizable operators
which couple to $R$ odd combinations of standard model particles.
This also ensures at least one LSP per decay.
The relic baryon and LSP number densities are then roughly equal.
The ratio of baryon to dark matter densities is then
naturally $\Omegab / \OmegaLSP \sim {\cal O}(\mb / \mLSP)$.
The resulting upper limit on the LSP mass is model
dependent but in the range
${\cal O}(30-140)$ GeV.
The total relic density is related to the order at which
the flat direction which gives rise to the condensate is lifted.
The observed density is obtained for a direction which is lifted
by a fourth order Planck scale suppressed operator in the superpotential.
}

\endpage

\parskip 0pt
\parindent 25pt
\overfullrule=0pt
\baselineskip=18pt
\tolerance 3500
\endpage
\pagenumber=1



\section{Introduction}

\REF\ewbaryo{For a review of electroweak baryogenesis see
A. Cohen, D. Kaplan, and A. Nelson, Ann. Rev. Nucl. Part. Sci.
{\bf 43} (1993) 27.}

\REF\ad{I. Affleck and M. Dine, Nucl. Phys. B {\bf 249} (1985) 361.
For a recent discussion of the Affleck-Dine mechanism of
baryogenesis see ref. \refmark{\drt}.}

\REF\decaybaryo{For a review of baryogenesis from the out of
equilibrium decay of heavy particles see
E. Kolb and M. Turner, Ann. Rev. Nucl. Part. Sci. {\bf 33} (1983) 645.}

The baryon asymmetry and dark matter density 
may provide indirect windows to very early epochs
in the evolution of the universe, and
to physics at large energy scales.
In most scenarios the physical mechanisms which give rise to
the baryon asymmetry and dark matter are unrelated.
For example, in supersymmetric theories the dark matter density
is usually assumed to result from the freeze out of the
lightest supersymmetric particle (LSP).
If $R$ parity is unbroken the LSP is stable, and the relic LSPs
make up the dark matter.
The baryon asymmetry is usually assumed to arise either
at the electroweak phase transition \refmark{\ewbaryo},
by the Affleck-Dine mechanism in which a coherent
condensate carrying baryon number is generated along
a supersymmetric flat direction \refmark{\ad},
or the out of equilibrium decay of massive particles
through baryon and $CP$ violating interactions \refmark{\decaybaryo}.
In all these mechanisms the dark matter and baryon
densities are a priori unrelated.
This is not surprising since the LSP carries a multiplicative
quantum number while baryon number is additive.
The processes which lead to the respective relic densities
are therefore distinct.
That the baryon and dark matter
densities are in fact the same within a few orders is not necessarily
a direct consequence of any of these mechanisms, and seems fortuitous.

\REF\techni{S. Nussinov, Phys. Lett. B {\bf 165} (1985) 55;
R. Chivukula and T. Walker, Nucl. Phys. B {\bf 329} (1990) 445;
S. Barr, R. Chivukula, and E. Farhi, Phys. Lett. B {\bf 241}
(1990) 387;
D. Kaplan, Phys. Rev. Lett. {\bf 68} (1992) 741.}

Here I suggest an alternate supersymmetric mechanism in
which {\it both} baryons and dark matter arise from the
late decay of a weak scale mass particle.
As discussed below if the mass of the decaying particle is above
the LSP mass, and the population of decaying particles carries
a large asymmetry, then (optimally) roughly equal
numbers of baryons and LSPs result
from the decay.
If the temperature at the era of decay is low enough,
the LSPs do not rethermalize, and the relic density is determined
by the decaying particle density.
The ratio of baryon to dark matter density in this scheme is
then proportional to the ratio of the lightest baryon mass
to LSP mass,
$\Omega_b / \Omegalsp \sim {\cal O} (\mb / \mlsp)$.
For an LSP with weak scale mass, this gives roughly the correct
ratio, $\Omega_b / \Omegalsp \sim {\cal O}(10^{-1}-10^{-2})$.
This result is reminiscent of the analogous relation in technicolor
theories if the lightest technibaryon makes up the dark matter.
There the electroweak anomaly ensures that at high
temperatures the
baryon and technibaryon number densities are roughly
equal \refmark{\techni}.
Here however, the LSP density is protected from erasure
by the low temperature
at the time of decay, rather than an additive quantum number.


\REF\lsgut{D. Lyth and E. Stewart, preprint Lancaster-TH-9502,
hep-ph/9502417.}

\REF\drt{M. Dine, L. Randall, and S. Thomas,
preprint SLAC-PUB-95-6776, hep-ph/9503303, to appear in
Phys. Rev. Lett.;
M. Dine, L. Randall, and S. Thomas,
preprint SLAC-PUB-95-6846.}

In order for this mechanism of relating the baryon and dark matter
densities to be operative 
the decay must occur below the LSP thermalization
temperature, but above the temperature at which nucleosynthesis
takes place.
This can happen if the decaying particle is coupled to standard
model fields by nonrenormalizable operators
suppressed by an intermediate scale, somewhat below the GUT
scale \refmark{\lsgut}.
These operators must carry baryon number if
any asymmetry is to result, and be odd under $R$ parity if at
least one LSP is to result from each decay.
In addition, there should be a large particle-antiparticle
asymmetry in the decaying
population if order one baryon per decay is to result.
Such large asymmetries can result from the coherent production
of scalar fields along supersymmetric flat directions.
Flat directions are likely to be generic features
of supersymmetric theories.
Finally, in order that the total density of the universe
have the observed value now, the number density of the late decaying
particles should be less than thermal at the time of decay.
Far too many LSPs would remain if the decaying particles had
thermal number density.
A subthermal number density in fact
naturally occurs for coherent production along flat
directions which are lifted by Planck scale suppressed terms in
the superpotential \refmark{\drt}.
The density in the condensate, and therefore the total relic density,
is related to the order at which the flat direction is lifted.
All the ingredients for this late decay
scenario therefore exist in supersymmetric theories.


\section{Requirements for Baryons and LSPs from Late Decay}


A number of requirements must be met if the late decay scenario
for the origin of the baryon asymmetry and dark matter is to
be realized within supersymmetry.
In most SUSY models the LSP is typically a neutralino, a linear
combination of gaugino and Higgsino.
If the relic LSPs are to act as dark matter, they must be stable as
the result of some symmetry.
Since the neutralino is Majorana,
this must be a discrete symmetry, giving a multiplicative
quantum number.
In what follows I will assume the required symmetry is
$R$ parity. 
If the decaying particle is much heavier than the LSP then
multiple LSPs can in principle be produced in the decay chain.
However, as discussed below at most one unit of baryon number
can result from each decay.
So unless the LSP is very light, there should not be too many
LSPs per decay.
In order to guarantee that at least one LSP results from each
decay, the decaying particle should be odd under $R$ parity
from the low energy point of view.
If the mass of the decaying particle is in the range
$\mlsp < \mphi < 2 \mlsp$, then precisely one LSP results per decay.
For simplicity this will be assumed to be the case.
The decaying particle then also has weak scale mass.

\REF\nucleoa{D. Lindley, Ap. J. {\bf 294} (1985) 1;
J. Ellis, D. Nanopoulos, and S. Sarkar,
Nucl. Phys. B {\bf 259} (1985) 175;
S. Dimopolous, R. Esmailzadeh, L. Hall, and G. Starkman,
Nucl. Phys. B {\bf 311} (1988) 699;
J. Ellis, G. Gelmini, J. Lopez, D. Nanopoulos, and S. Sarkar,
Nucl. Phys. B {\bf 373} (1992) 399.}

\REF\nucleob{M. Reno and D. Seckel, Phys. Rev. D {\bf 37} (1988) 3441;
G. Lazarides, R. Schaefer, D. Seckel, and Q. Shafi,
Nucl. Phys. B {\bf 346} (1990) 193.}


\REF\gravitino{J. Ellis, A. Linde, and D. Nanopoulos, Phys. Lett.
B {\bf 118} (1982) 59;
D. Nanopoulos, K. Olive, and M. Srednicki, Phys. Lett. B
{\bf 127} (1983) 30;
M.Yu. Khlopov and A.D. Linde,
Phys. Lett. {\bf 138B} (1984) 265; J. Ellis, J.E. Kim
and D.V. Nanopoulos, Phys. Lett. {\bf 145B} (1984) 181.}

\REF\polonyi{C. Coughlan, W. Fischler, E. Kolb, S. Raby,
and G. Ross, Phys. Lett. B {\bf 131} (1983) 59; J. Ellis,
D.V. Nanopoulos and M. Quiros, Phys. Lett. B {\bf 174}
(1986) 176.}

If the number density of relic LSPs is to be determined by the
density of decaying particles, the temperature during the
decay epoch should be less than the LSP equilibration temperature.
If the decay takes place above this temperature, the relic
LSP density is determined by freeze out, as in the usual
scenario.
For particles with weak scale
annihilation cross section and mass, the equilibration
temperature is roughly $T \sim {1 \over 20} m_{\rm LSP}$.
With the LSP mass in the range discussed below this corresponds to
roughly $T \sim {\cal O}(1~{\rm GeV})$.
In addition to this upper limit on the temperature at the
time of decay there is a lower limit arising from
nucleosynthesis.
If decays take place during or after nucleosynthesis the light
element abundances can be modified by photodissociation
and photoproduction by decay products \refmark{\nucleoa,\nucleob}.
This can be avoided for $T \gsim $ 1 MeV since the weak interactions
are in equilibrium and the usual neutron to proton ratio results.
The decay temperature must therefore lie in the window
1 MeV $\lsim T \lsim$ 1 GeV.
The decay rate, $\Gamma$, and decay temperature, $T_d$ are related by
$T_d^2 \sim \sqrt{90/g_* \pi^2} \Gamma M_p$,
where $M_p=m_p / \sqrt{8 \pi}$
is the reduced Planck mass, and $g_*$ is the effective number
of degrees of freedom ($g_* \simeq 10.75$ for $T \sim$ 1 MeV).
With weak scale mass, such a slow decay rate
implies the decaying particle must couple to standard model fields
only through nonrenormalizable interactions.
Decay through Planck scale suppressed couplings leads to a decay
temperature much too low to avoid the bounds from
nucleosynthesis \refmark{\gravitino,\polonyi}.
However, a decay temperature of order the nucleosynthesis bound
in fact results if the particle decays through dimension 5 operators
suppressed by a scale somewhat below the GUT scale \refmark{\lsgut}.
For the 3-body decays discussed below
$$
\Gamma \simeq { 6 \bar{\lambda}^2 \mphi^3 \over (8 \pi)^3 M^2}
\eqn\rate
$$
where
$\bar{\lambda}^2=\sum |\lambda|^2$ is a sum over
generations in the final state,
$\lambda / M$ is the coefficient of the operator,
$\mphi$ is the mass of the decaying particle,
and final state masses have been neglected for simplicity.
This gives a decay temperature of
$$
T_d \sim ~.3 ~\left( {10^{14}~{\rm GeV} \over M / \bar{\lambda} } \right)
\left( { \mphi \over 100 ~{\rm GeV} } \right)^{3/2} ~ {\rm MeV}
\eqn\decaytemp
$$
A decay temperature in the window given above
can be obtained for
$3 \times 10^{10}$ GeV $\lsim M/\bar{\lambda} \lsim
3 \times 10^{13}$ GeV.
Although this is probably too low to be associated directly with
the GUT scale, it could arise from an intermediate scale.

Producing a baryon asymmetry in the decay
imposes a number of additional requirements.
The particle must of course decay through an operator which
transforms under $U(1)_B$ with respect to the standard model fields.
In principle nonrenormalizable
couplings could arise from $D$ type Kahler potential
terms or $F$ type superpotential terms.
However, with conserved $R$ parity, the gauge invariant operators
which carry baryon number contain at least 3 standard model fields.
A Kahler potential coupling of this type
to the decaying particle is
dimension 6, but a superpotential coupling to 3 fields
is dimension 5.
The only invariant made out of 3 standard model fields
which carries $U(1)_B$ is $\bu \bd \bd$.
The unique superpotential coupling which satisfies the requirements
is therefore
$$
W= {\lambda \over M} ~\phi \bu \bd \bd
\eqn\Wcoupling
$$
where $\phi$ is the decaying particle and
generation indices are suppressed.
Notice that  $\bu \bd \bd$ is odd under $R$ parity.  So with an unbroken
$R$ parity (at least) one LSP results from each decay.
In addition, if $R$ parity is to remain unbroken after the decay,
$\phi=0$ must be the ground state.

Depending on the specific model $\phi$ might decay through
other dimension 5 terms in addition to
\Wcoupling.
Decay through superpotential couplings to the other
$R$ odd combinations of 3 standard model
fields, namely $L \bd Q$, and $LL\be$ (which do not carry baryon
number) would still give at least one LSP
per decay, but dilutes the baryon number (for the decay
from a condensate with a particle-antiparticle asymmetry
discussed below).
In GUT theories, such operators are in general related by GUT
symmetries.
For example, in $SU(5)$ models $\bu \bd \bd$, $L \bd Q$ and
$LL \be$ are contained in $\bar{5} \bar{5} 10$.
The existence of these other decay channels related by
$SU(5)$ would dilute the baryon number by a factor ${3 \over 7}$.
All other dimension 5 couplings are through operators which
do not carry baryon number.
These couplings include: 1)
superpotential couplings to $R$ even
combinations of standard model fields, namely
$Q H_u \bar{u}$, $Q H_d \bar{d}$, and $L H_d \be$,
2) Kahler potential couplings
$$
{\lambda^{\prime} \over M} \phi \chi^{\dagger} \chi
\eqn\kahlercoupling
$$
where $\chi$ is a light field,
and 3) $\phi$
dependence of the gauge kinetic functions 
$$
{g^2 \over 32 \pi^2 M} ~\phi W^{\alpha} W_{\alpha}
\eqn\phiWW
$$
where $W_{\alpha}$ is the field strength for a light
gauge supermultiplet.
All of these decay modes of course do not contribute to the
baryon asymmetry.
However, if $\mphi < 2 \mlsp$ no LSPs result either.
So if $\phi$ is light enough these decay modes do not affect the
relic $\Omega_b / \Omega_{\rm LSP}$.
Finally, the 
coupling $\phi L H_u$, if present, would allow decay through
a renormalizable operator, giving a very large decay temperature.
In addition, it would cause $\phi$ to pair
up with some linear combination of neutrinos after electroweak
symmetry breaking, giving a Dirac neutrino with weak scale mass.
This (dangerous) coupling must therefore be restricted
in some way (as in the toy model given in the next section).

\REF\bbaryon{D. Nanopoulos and S. Weinberg, Phys. Rev. D {\bf 20}
(1979) 2484.}

\REF\rbaryon{
A baryon symmetric but $CP$ asymmetric primary decay can be
transformed into a
non-negligible baryon asymmetry in the decay chain by
renormalizable baryon violating interactions.
However, with minimal low energy particle content this requires
$R$ parity violation.
S. Dimopoulos and L. Hall, Phys. Lett. B {\bf 196} (1987) 135;
J. Cline and S. Raby, Phys. Rev. D {\bf 43} (1991) 1781;
S. Mollerach and E. Roulet, Phys. Lett. B {\bf 281} (1992) 303.
}

The decay through the operator \Wcoupling~can in principle
lead to a net baryon asymmetry, parameterized
by $\epsilon = \langle N_b \rangle / \langle N_{\rm LSP} \rangle$,
where $\langle N_b \rangle$ and
$\langle N_{\rm LSP} \rangle$ are the average number of baryons
and LSPs resulting from each decay.
In order for $\Omega_b / \OmegaLSP \sim {\cal O}(m_b / \mLSP)$
to hold, $\epsilon$ should not be too small.
Direct production of a baryon asymmetry in the decay requires
decay channels which carry different baryon number,
final state interactions, and
$CP$ violating interference terms which contain at least
two baryon violating couplings \refmark{\bbaryon}.
With conserved $R$ parity, baryon number is violated only
by nonrenormalizable operators, giving negligible interference terms.
Any baryon asymmetry produced directly
in the decay is therefore insignificant \refmark{\rbaryon}.
However, a nonzero $\langle N_b \rangle$
will be transferred to baryons through the operator
\Wcoupling~if there is an initial
particle-antiparticle asymmetry in the population of decaying
particles.
Since $\epsilon$ should not be too small, there must
exist a near maximal asymmetry in the decaying population.

\REF\nrt{M. Grisaru, W. Siegal, and M. Rocek, Nucl. Phys. B
{\bf 159} (1979) 429;
N. Seiberg, Phys. Lett. B {\bf 318} (1993) 469. }

Such a large asymmetry might appear hard to achieve.
However, the coherent production of a scalar condensate along
a supersymmetric flat direction can give rise to a large
asymmetry in the condensate \refmark{\ad,\drt}.
Here, flat direction refers to a direction in field space
on which the perturbative potential vanishes at the
renormalizable level.
Such directions are generic in supersymmetric theories.
The nonrenormalization theorem protects these directions
from being lifted by quantum corrections \refmark{\nrt}.
In the presence of SUSY breaking, a potential can arise though.
Whether or not a condensate is actually generated along a flat direction
depends on the sign of the SUSY breaking soft mass term at early times,
$m^2 \phi^* \phi$, where $\phi$ parameterizes the flat direction.
When $H \gsim \mgravitino$ the finite energy density
of the universe induces soft parameters
along flat directions with a scale set by the Hubble
constant \refmark{\drt}.
If the induced $m^2 > 0$, the origin is stable, and the large
expectation values required to form a condensate do not
arise.
However, if the induced $m^2 < 0$, the origin
is unstable and large expectation values can develop.
In this case, if the flat direction is lifted at order $n$
by a nonrenormalizable
operator in the superpotential
$$
W = { \beta \over n M^{n-3}_n} \phi^n
\eqn\Wn
$$
then the relevant part of the potential along the flat direction is
$$
V(\phi) = (c H^2 + m_{\phi}^2) |\phi|^2
 + \left( {(A + a H) \beta \phi^n \over n M_n^{n-3}} ~ + h.c. \right)
 + | \beta|^2 {|\phi|^{2n-2} \over M_n^{2n-6} }
\eqn\vphi
$$
where $m_{\phi} \sim A \sim \mgravitino$ are soft
parameters arising from hidden sector SUSY breaking, and
$c \sim a \sim {\cal O}(1)$ are the soft parameters induced
by the finite energy density \refmark{\drt}.
Here the scale $M_n$ may in general be (much) different
than the scale of the operators which allow $\phi$ to decay.
For $c < 0$ the expectation value along the flat direction is
determined at early times by a balance between the mass term
and nonrenormalizable terms.
If $m_{\phi}^2 > 0$, then when $H \sim \mgravitino$ the origin
becomes stable and the field begins to oscillate freely with a
large initial value.
However, at just this time since the expectation value
of the field is determined by a balance between the mass and
nonrenormalizable terms, the $U(1)$ violating $A$ term necessarily
has the same magnitude.
Depending on the initial phase of the field, the presence of
the $A$ term with this magnitude can lead to a near maximal asymmetry
in the condensate.
So if a condensate is produced along a flat direction which is lifted by
a nonrenormalizable superpotential, it naturally has a large
asymmetry.
For a flat direction made of squark or slepton fields, this
is the mechanism of baryogenesis proposed by
Affleck and Dine \refmark{\ad}.
Here however, the initial condensate asymmetry is in the
$\phi$ field, and is only
transferred to baryons by decay through the operator \Wcoupling.

\REF\lsp{The relic LSP density arising from gravitino decay
has been considered by L. Krauss, Nucl. Phys. B {\bf 227} (1983) 556;
and from axino decay by K. Rajagopal, M. Turner, and F. Wilczek,
Nucl. Phys. B {\bf 358} (1991) 447.}

The final, and perhaps most nontrivial requirement, is that
$\Omegalsp + \Omega_b \simeq 1$.
If the decaying particles dominate the energy density
at the time of decay,
the universe is in a matter dominated era at that epoch.
However, since $\mlsp$ and $\mphi$ are the same order this would imply
the universe remained matter dominated below this temperature.
Matter domination from such an early epoch is incompatible with
nucleosynthesis \refmark{\nucleoa}.
While it may have been natural for the condensate to dominate
the energy density, this is clearly unacceptable \refmark{\lsp}.
The condensate must have a small enough energy density so
that matter domination from the relic LSPs starts at a
temperature of $T \sim 5~\Omega h^2$ eV, where
$h = H/(100~{\rm km}~s^{-1}~{\rm Mpc}^{-1})$, and $H$ is the Hubble
constant now.
Assuming critical density, $\nlsp=1$ in the decay, and given the current
temperature,
the condensate number density at the time of decay
can be parameterized as
$$
{ \nphi \over s} \simeq 7 \times 10^{-11} ~h^2~
 \left( {50~{\rm GeV} \over \mlsp } \right)
\eqn\ns
$$
where $s=(2 \pi^2 g_*/45) T^3$ is the entropy density at the time of decay.
This is much less than a thermal number density, $n/s \sim 1/g_*$.
Now the total density in the condensate is determined by the expectation
value when the field begins to oscillate freely.
{}From \vphi~the value of the field when oscillations begin
($H \sim \mgravitino$) is
$$
\phi_0 \simeq \left( {\alpha \mphi M_n^{n-3} \over \beta}
\right)^{1 \over n-2}
\eqn\phinot
$$
where $\alpha$ is a constant of order unity.
The fractional energy density in the condensate when
oscillations begin is
$\rho_{\phi} / \rho_{\rm tot} \simeq \phi_0^2 / M_p^2 \ll 1$.
In an inflationary scenario with a reheat temperature low
enough to avoid overproducing gravitinos by thermal rescatterings,
the universe is in an inflaton matter dominated era when
$H \sim \mgravitino$ \refmark{\drt}.
So $\rho_{\phi} / \rho_{\rm tot}$ stays roughly constant until the inflaton
decays.
After the inflaton decays the condensate density per entropy
density is
$$
{\nphi \over s} \sim {T_R \over \mphi M_p^2}
  \left( {\mphi M^{n-3} \over \beta} \right)^{ 2 \over n-2}
\eqn\nsphi
$$
where $T_R$ is the reheat temperature after inflation.
Without additional entropy releases
$n_{\phi}/s $ stays constant until the time of decay.
So the relic fractional density in the condensate is determined
by the order at which the flat direction is lifted, and the
reheat temperature after inflation.
For $M_n \sim M_p$ and $n \geq 6$ this is generally too large
for reasonable reheat temperatures.
However for $n=4$ and $M_n / \beta \sim M_p$ ,
$\nphi / s \sim T_R / M_p$.
With $T_R \sim 10^8$ GeV, this gives just the
required condensate density.
Therefore if the late decaying condensate arises from
coherent production along a direction which is lifted by a
Planck suppressed
fourth order term in the superpotential, the required relic
density naturally arises for reasonable values of the
reheat temperature after inflation.

In addition to the required fourth order Planck suppressed
term in the superpotential,
there are in general higher order SUSY breaking terms
(in addition to the mass term) which
are suppressed by the Planck scale.
Assuming hidden sector SUSY breaking
these give a general form for the soft potential of
$$
V_s(\phi) = \mgravitino^2 M_p^2 {\cal F}(\phi / M_p)
\eqn\vsoftplanck
$$
However, just on energetic grounds the nonrenormalizable
term in the superpotential forces $\phi \ll M_p$.
The higher order corrections in \vsoftplanck~are therefore
unimportant.
In addition, there are higher order soft terms generated
by integrating out fields which gain mass at the scale $M$.
These are of the general form
$$
V_s(\phi)= \mgravitino^2 M^2 {\cal G}(\phi/M)
\eqn\vsoftM
$$
With $M$ in the range required to give an acceptable decay
temperature for $\phi$, these higher order terms are
less important for $H \sim \mgravitino$ than the terms in
\vphi~with $n=4$.
The higher order SUSY breaking potential terms for $\phi$
therefore do not spoil the expectation that the condensate
carries a large asymmetry, or the prediction for the
relic density.




\section{A Toy Model for Baryons and LSPs from Late Decay}

\REF\rnote{Under a discrete $Z_N$ $R$ transformation,
a chiral superfield transforms as
$\Phi \to e^{2 \pi i R/N} \Phi$, where $R$ is the $Z_N$
$R$ charge of the field.
The scalar, fermionic, and auxiliary  components transform as
$A \to e^{2 \pi i R/N} A$,
$\psi \to e^{2 \pi i (R-1)/N} \psi$, and
$F \to e^{2 \pi i (R-2)/N} F$.
The superpotential transforms as $W \to e^{4 \pi i /N} W$.}


It is easy to build models which satisfy all the requirements
outlined in the previous section.
As an existence proof, consider the following toy model.
The flat direction required for the coherent production
can be parameterized by a singlet field $\phi$.
In principle this could be a composite field in some sector
of the theory,
but here will be taken to be an elementary singlet for simplicity.
The singlet $\phi$ should be protected from obtaining a large mass
while allowing the operator \Wcoupling.
This can be enforced with discrete symmetries.
For example, under a $Z_4$ discrete $R$ symmetry the superpotential
transforms as $W \rightarrow -W$ \refmark{\rnote}.
If $\phi$ and all the  $\bu$ and $\bd$ transform as
$f \rightarrow e^{i \pi /4} f$, where
$f \in ( \phi, \bu, \bd)$,
then the
operator \Wcoupling~is allowed while a superpotential mass term,
$m \phi \phi $, is not allowed.
The usual Yukawa couplings,
$\lambda_u Q H_u \bu$,
$\lambda_d Q H_d \bd$, and
$\lambda_e L H_d \be$, are allowed
if the other standard model fields transform under the $Z_4$ as
$L \rightarrow L$,
$Q \rightarrow e^{i \pi /4} Q$,
$h \rightarrow e^{i \pi /2} h$, where
$h \in (H_u, H_d, \be)$.
The operator \Wcoupling~must be generated by integrating out
particles with intermediate scale mass.
This can be accomplished in this model by introducing a Dirac pair
$U^{\prime}$ and $\bar{U}^{\prime}$, with mass,
$m_U U^{\prime} \bar{U}^{\prime}$, and Yukawa couplings
$g \bar{U}^{\prime} \bd \bd$, and $g_{\phi} \phi \bu U^{\prime}$.
These couplings and Dirac mass can be enforced by the
transformation $U^{\prime} \rightarrow e^{i \pi /2} U^{\prime}$ and
$\bar{U}^{\prime} \rightarrow e^{i \pi /2} \bar{U}^{\prime}$.
The mass scale $m_U$ could arise from dynamics which
preserves the discrete symmetry.
Integrating out the Dirac pair gives the operator \Wcoupling~with
$\lambda / M = g g_{\phi}/m_U$.
The operator \phiWW~is also generated at the scale $M$, but
is suppressed by a loop factor compared with \Wcoupling.
So in this model the dominant decay mode is
$\phi \rightarrow \bu \bd \bd$.
Finally,
the dangerous superpotential coupling $\phi L H_u$ is restricted
by the discrete symmetry.


The flat direction $\phi$ can be lifted by nonrenormalizable terms in
the superpotential.
With the $Z_4$ $R$ symmetry, the lowest order term in the
superpotential is $\phi^4$.
Such an operator is not generated at the intermediate
scale, but presumably can arise directly at the Planck scale
$$
W={ \beta \over M_p} \phi^4
\eqn\phifoursuper
$$
And, as discussed in the last section,
an operator lifted at fourth order in the superpotential and
suppressed by the Planck scale is precisely what is required to
give the correct magnitude for the dark matter and baryon densities.
In addition to the fourth order term in the superpotential,
$\phi$ is lifted by a
fourth order SUSY breaking term in the soft potential
generated by integrating out
the heavy Dirac pair $U^{\prime} \bar{U}^{\prime}$
$$
V_s(\phi) \simeq
{ g_{\phi}^4 \mgravitino^2 \over 16 \pi^2 m_U^2} (\phi^* \phi)^2
\eqn\phifoursoft
$$
However, as discussed in the previous section, for
$H \sim \mgravitino$, terms of this order are subdominant
compared with \phifoursuper.

Acceptable soft SUSY breaking terms can also result in this model.
In order to allow visible sector gaugino masses the $Z_4$ $R$ symmetry
must be broken in the SUSY breaking sector to $Z_2$ $R$ parity.
For definiteness consider a hidden sector scenario in which
SUSY breaking is transmitted by Planck suppressed interactions.
The breaking to $Z_2$ $R$ parity
can be accomplished with a hidden sector field $z$
which is invariant under $Z_4$ and
breaks SUSY by an auxiliary component expectation value
$\langle F_z \rangle \sim \sqrt{\mgravitino M_p}$.
In addition, the soft $A$ term $A \beta \phi^4/M_p$,
required to generate an asymmetry in the $\phi$ condensate,
can arise from supergravity interactions, and
the Kahler potential coupling
${1 \over M_p}  \intD z \phi^{\dagger} \phi$.
Dimension 3 soft $A$ terms for standard model fields
arise from similar couplings.
A soft $H_u H_d$ scalar mass and weak scale $\mu$ term can arise from
Kahler potential couplings
${1 \over M_p^2} \intD z^{\dagger} z^{\prime} H_u H_d$ and
${1 \over M_p} \intD z^{\prime} H_u H_d$, where
$z^{\prime}$ is a hidden sector field which participates in
SUSY breaking and
transforms as $z^{\prime} \to - z^{\prime}$ under the discrete symmetry.
A weak scale mass for the flat direction,
$\mgravitino^2 \phi^* \phi$, results from supergravity interactions
and/or Kahler potential couplings with
the hidden sector.
However, most importantly, with the hidden sector couplings
sketched above,
the $Z_4$ symmetry does not allow a soft mass term
$\mgravitino^2 \phi \phi$ from Kahler potential couplings,
which would violate the $U(1)$ carried by $\phi$.
The classical evolution of the condensate
at late times
therefore preserves the asymmetry
generated when the coherent oscillations begin.

So in this model all the requirements are satisfied with a single
discrete symmetry.
Although the model is perhaps unrealistically simple, it demonstrates
that all the requirements for baryons and LSPs from late decay
of a condensate can be met in a technically natural manner.


\section{Conclusions}

\REF\nucleo{T. Walker, G. Steigman, D. Schramm, K. Olive, and H. Kang,
Astrophys. J. {\bf 376} (1991) 51;
W. Smith, L. Kawano, and R. Malaney, Astrophys. J. Suppl.
{\bf 85} (1993) 219;
For recent discussions of the limits on $\omegab h^2$ from
nucleosynthesis see
P. Kernan and L. Krauss, Phys. Rev. Lett. {\bf 72} (1994) 3309;
K. Olive and G. Steigman, preprint UMN-TH-1230/94, OSU-TA-6/94,
astro-ph/9405022.}

In the late decay scenario outlined here,
the baryon and LSP densities are related by
$$
{\Omega_b \over \OmegaLSP} \simeq \epsilon~{\mb \over \mlsp}
\eqn\rhorho
$$
As discussed above, $\epsilon < 1$
if one LSP results from each decay.
Under the assumption that the total density is near critical,
$\omegab + \omegalsp \simeq 1$, a lower limit on the baryon density
then gives an upper limit on the LSP mass in this scheme.
For $\Omega_b \ll \omegalsp$,
$$
\mlsp \simeq \epsilon~ {\mb \over \omegab}
$$
The absolute lower bound on $\omegab$ comes from the observed
density of luminous matter, $\omegab \gsim .007$.
This gives an upper limit of $\mlsp \lsim 140~\epsilon$ GeV.
A more stringent upper limit comes from nucleosynthesis.
The primordial light element abundances 
depend on the baryon to entropy ratio at the time of nucleosynthesis.
Comparison of the calculated and observed abundances
gives upper and lower limits on the baryon density,
$.01 \lsim  \omegab h^2 \lsim .015$ \refmark{\nucleo}.
The nucleosynthesis lower limit on the baryon density
gives the upper limit $\mlsp \lsim 100h^2 ~\epsilon$ GeV.
So 
the LSP is expected to be fairly light in this late decay scenario.
For example, in $SU(5)$ models for which $\epsilon < {3 \over 7}$, with
a hubble constant $h < .8$, the upper limit on the LSP mass
is $\mlsp < 30$ GeV.

\REF\mixed{J. Ellis, D. Nanopoulos, L. Roszkowski, and D. Schramm,
Phys. Lett. B {\bf 245} (1990) 251.}

\REF\mixlight{L. Roszkowski, Phys. Lett. B {\bf 262} (1991) 59;
L. Roszkowski, Phys. Lett. B {\bf 278} (1992) 147;
J. McDonald, K. Olive, and M. Srednicki, Phys. Lett. B {\bf 283}
(1992) 80;
A. Bottino, V. De Alfaro, N. Fornengo, G. Mignola, and M. Pignone,
Astropart. Phys. {\bf 2} (1994) 67.}

\REF\coann{K. Griest and D. Seckel, Phys. Rev. D {\bf 43} (1991) 3191;
S. Mizuta and M. Yamaguchi, Phys. Lett. B {\bf 298} (1993) 120;
M. Dress and M. Nojiri, Phys. Rev. D {\bf 47} (1993) 376.}

In addition to the LSPs arising from the late decaying condensate
there will be a population of LSPs arising from thermal freeze out.
However, 
the low mass required for the late decay scenario
can give a freeze out density which is well below critical.
For an LSP which has a sizeable mixture of Higgsino and
gaugino components,
annihilation through $s$-channel $Z$ exchange is very efficient and
leaves a very small relic density from freeze
out \refmark{\mixed,\mixlight}.
For a mostly Higgsino LSP, coannihilation with the
other 
Higgsino states also leads to negligible relic density \refmark{\coann}.
A small relic density for a light nearly pure gaugino LSP
can also result from annihilation through
$t$-channel squark and slepton exchange if one of the sleptons
or squarks are light \refmark{\mixlight}.
So depending on the precise composition of the LSP, the
late decay can give the dominant contribution to the relic
LSP density.
Independent of the production of a baryon asymmetry, late decay
is an interesting source of relic LSPs in the low mass
regime.
In fact, if the LSP was found to be in a region of parameter space
for which the freeze out density was too small to give closure
(such as the light Higgsino or mixed Higgsino-gaugino regions)
the only alternate source for LSPs would be a late decay {\it below}
the freeze out temperature.



In conclusion, the ratio of dark matter to baryon density can
naturally be ${\cal O}(10-100)$
if stable weak scale mass particles and baryons result
in roughly equal amounts
from the late decay of a particle.
A natural way in which this can occur is for a condensate
with a large net asymmetry to decay to $R$ odd combinations
of standard model fields.
Supersymmetric theories with a conserved $R$ parity can in
principle have all the ingredients to realize this scenario.

I would like to thank R. Brandenberger,
M. Dine, J. Primack, L. Randall, and U. Sarid for useful
discussions, and J. Frieman, R. Malaney, and G. Steigman
for discussions about nucleosynthesis.

\refout
\bye